\documentclass[stackrel,amsmath,amssymb,aps,pre,twocolumn,showpacs,floatfix]{revtex4}
\usepackage{graphicx}
\usepackage{dcolumn}
\usepackage{bm}
\usepackage{color}

\begin{document}

\title{Efficiency fluctuations in quantum thermoelectric devices}

\author{Massimiliano Esposito}
\affiliation{Complex Systems and Statistical Mechanics, Physics and Material Science Research Unit, University of Luxembourg, Luxembourg}
\author{Maicol A. Ochoa}
\affiliation{Department of Chemistry \& Biochemistry, University of California San Diego, La Jolla CA 92093, USA}
\author{Michael Galperin}
\affiliation{Department of Chemistry \& Biochemistry, University of California San Diego, La Jolla CA 92093, USA}

\date{\today}

\begin{abstract}
We present a method, based on characterizing efficiency fluctuations, 
to asses the performance of nanoscale thermoelectric junctions.
This method accounts for effects typically arising in small junctions, 
namely, stochasticity in the junction's performance, quantum effects, 
and nonequilibrium features preventing a linear response analysis.  
It is based on a nonequilibrium Green's function (NEGF) approach, which 
we use to derive the full counting statistics (FCS) for heat and work, and which in 
turn allows us to calculate the statistical properties of efficiency fluctuations.
We simulate the latter for a variety of simple models where our method is exact.
By analyzing the discrepancies with the semi-classical prediction of a quantum master 
equation (QME) approach, we emphasize the quantum nature of efficiency fluctuations 
for realistic junction parameters.  
We finally propose an approximate Gaussian method to express efficiency 
fluctuations in terms of nonequilibrium currents and noises which are 
experimentally measurable in molecular junctions. 
\end{abstract}


\maketitle

\section{Introduction}

The development of thermoelectric materials is at the forefront of the research related 
to energy conversion and storage. While research on thermoelectricity in bulk materials 
goes back to the middle of the last century~\cite{Ioffe_1957}, measurements at the 
nanoscale (and in particular, studies of thermoelectricity in molecular junctions) 
were only reported recently~\cite{MajumdarMcEuenPRL01}.  
The small size of the junctions gives rise to new physical phenomena, not accessible at the 
macroscopic level, and which are considered promising for reaching more effective energy conversion. 
The thermoelectric properties of nanoscale junctions have indeed received a lot of attention
in the last years, both experimentally~\cite{MajumdarScience07, ParkNatMat07, MajumdarCPL10, VenkataramanNL12, CahiullNatMater12, PhysRevLett.109.016601, TsutsuiSciRep13, CuevasReddyNature13, EsslingerScience13, ChabinycNatMater14, ReddyNatNano14, LeePRL14} and theoretically~\cite{LakeDattaPRB92, PaulssonDattaPRB03, LinkePRL05, GalperinNitzanRatnerMP08, LiuChenACSNano09, LiuChenPRB09, DubiDiVentraNL09, WegewijsFlensbergPRB10, EntinWohlmanImryAharonyPRB10, FranssonGalperinPCCP11, LiuChenJPCC11, DiVentraRMP11, EntinWohlmanAharonyPRB12, ThygesenJCompEl12, SanchezLopezPRL13}.

Experimental studies on thermoelectricity in nanoscale junctions make use of 
the macroscopic theory of thermoelectricity to asses the junction's performance. 
The latter is characterized by the figure of merit, a quantity exclusively 
defined in terms of linear response transport coefficients and thus 
ill-defined out of nonequilibrium. While the linear theory is reasonable in 
bulk material, it fails in small thermoelectric junctions which can operate 
in the nonlinear regime (for instance in the resonant tunneling regime). 
This fact motivated a number of studies to consider the macroscopic 
efficiency of the junction as an alternative to the figure 
of merit to characterizes the performance of the junction
\cite{EspositoLindenbergVanDenBroeckPRL09, EspositoPRB09, EspoLindVdB_EPL09_Dot, 
Esposito2010d, EspoKumLindVdBPRE12, LinkePRB10, VandenBroeckPRL12, LutzPRL12, 
ThessPRL13, BrandnerSaitoSeifertPRL13, SeifertBrandnerNJP13, WhitneyPRL14, LutzPRL14,vonOppenPRB14}.
The macroscopic efficiency is the traditional thermodynamic efficiency 
of a heat engine defined as the fraction of average power output extracted 
from the heat arising from the hot source. It is well defined far 
from equilibrium and upper bounded by the Carnot efficiency.

The nonequilibrium features of the junction are not the only characteristic 
to be accounted for at the nanoscale. Due to the small size of the system,
thermal fluctuations will play a much more import role than in bulk 
samples, resulting in a high variability in the junction's performance. 
This variability requires a statistical characterization of the energy 
conversion which can be performed using the methods of stochastic 
thermodynamics \cite{JarzynskiRev11, Seifert12Rev, EspVDBRev2014, Sekimoto10}. 
Such studies have been recently done for small classical energy converters
~\cite{EspositoVanDenBroeckNatCommun14, VanDenBroeckEspositoPRE14, 
EspoPoleVerl14, GeisslerNJP14, Jayannavar14, VanDenBroeckProesmans14}. 
The main idea is to define the efficiency along a single realization of 
the operating device and to develop techniques to study its fluctuations.
Experimental studies of efficiency fluctuation have been very recently 
performed in Ref.~\cite{ParrondoRoldan14}.

The third central feature of small thermoelectric junctions which needs
to be accounted for are quantum effects. Indeed, quantum coherences 
can significantly affect charge and energy transfers in molecular 
junctions as discussed theoretically in Refs~\cite{SanvitoPRB08, WackerPRB11, 
WhiteFainbergGalperinJPCL12, PeskinGalperinJCP12, GalperinNitzanJPCB13, WhitePeskinMGPRB13}
and shown experimentally in Refs.~\cite{JoachimInorgChem97, MayorWeberAngChim03, 
FlemingScience07, EngelFlemingNature07, ThossPRL11, VenkataramanHybertsenNatNano12}.

In this paper we provide a general method to study the performance of 
nanoscale thermoelectric junctions based on efficiency fluctuations.  
This methods accounts for the three key features characterizing small 
junctions, namely, the variability in performance due to fluctuations, 
operation modes arbitrary far from equilibrium, and quantum effects.   
It is based on the joint energy and particle full counting statistics (FCS) 
which we calculate within the nonequilibrium Green's functions (NEGF) formalism. 
While the quantum FCS of particle currents (e.g. electrons) in junctions 
is well developed~\cite{Levitov1996, UtsumiSchonPRL06, GogolinKomnikPRB06, 
SchonhammerPRB07, EspositoRMP09, SchonhammerJPCM09, EmaryPRB09, 
UtsumiSchonPRB10, FranssonGalperinPRB10, EspGalpJPCC10, 
ParkGalperin_FCS_PRB11, UtsumiPRB13}, that of energy was 
mostly limited to the quantum master equation (QME) approach
~\cite{EspositoRMP09, EspositoLindenberg, ButtikerSanchez12EPL, 
ButtikerSanchez11PRL, SegalPCCP12, BrandesEspositoNJP13, SchnirmalUtsumiPRB14} 
with its known limitations~\cite{EspGalpJPCC10, LeijnseWegewijsPRB08, EspGalpPRB09}.
By numerically calculating efficiency fluctuations for a set of simple models 
and comparing our NEGF results with those obtained using a QME approach, we 
identify the regimes where efficiency fluctuations display truly quantum features.
Moreover, we propose an approximate Gaussian scheme enabling to estimate efficiency 
fluctuations solely based on experimentally measurable quantities in molecular junctions 
~\cite{RuitenbeekNL06, TalRuitenbeekPRL08, Tal2008, YeyatiRutenbeekPRL12, BerndtPRL12, NatelsonPRB12}, 
namely the nonequilibrium energy and matter currents and noises.

The structure of the paper is the following. 
After introducing the FCS of energy, work, and heat within NEGF in Section~\ref{FCS}, 
we consider efficiency fluctuations in Section~\ref{efffluct}. 
In Section~\ref{numres}, we numerically evaluate efficiency fluctuations 
for various models, compare our results with the QME approach and 
describe the approximate scheme to estimate efficiency fluctuations experimentally. 
We summarize our findings in section~\ref{conclude}.

\section{FCS of particle and energy fluxes}\label{FCS}

The particle FCS for a single level strongly coupled to Fermi reservoirs was 
derived in Ref.~\cite{GogolinKomnikPRB06} and generalized to a multilevel
interacting system in Ref.~\cite{FranssonGalperinPRB10}.
Later, the methodology was applied to describe inelastic transport
in junctions in Ref.~\cite{ParkGalperin_FCS_PRB11}, where
the role of quantum coherence on the FCS was discussed.

\begin{figure}[t]
\centering\includegraphics[width=0.8\linewidth]{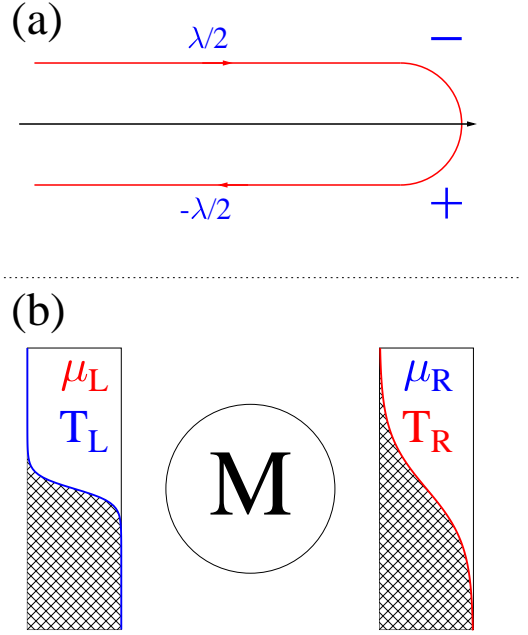}
\caption{\label{fig1}
(Color online) (a) Sketch of the counting field $\lambda$ dressing of 
the Keldysh contour forward ($-$) and backward ($+$) branches.
(b) Sketch of the a nano-thermoelectric junction consisting of a molecule $M$
embedded between two contacts $L$ and $R$ with $T_L<T_R$ and $\mu_L>\mu_R$.
}
\end{figure}

Here we extend the methodology to count particles and 
energy in a system strongly coupled to its reservoirs. 
Similar to the particle FCS~\cite{EspositoRMP09}, the treatment starts by 
dressing the evolution operator, $\hat U(t,t')$ with particle, $\gamma_K^P$, 
and energy, $\gamma_K^E$, counting fields at interface $K$ of the junction
\begin{equation}
 \hat U_\gamma(t,t') = e^{-i\gamma_K^P\hat N_K} e^{-i\gamma_K^E\hat H_K}
 \hat U(t,t') e^{+i\gamma_K^P\hat N_K} e^{+i\gamma_K^E\hat H_K}
\end{equation}
Note that $[\hat N_K;\hat H_K]=0$. The counting fields depend on the Keldysh contour branch
(see Fig.~\ref{fig1}a)
\begin{equation}
\gamma_K=
\begin{cases}
+\lambda_K^P/2 & \mbox{at } -  \\
-\lambda_K^P/2 & \mbox{at } +
\end{cases}
\qquad
\gamma_K^E=
\begin{cases}
+\lambda_K^E/2 & \mbox{at } -  \\
-\lambda_K^E/2 & \mbox{at } +
\end{cases}
\end{equation}
Here $-$ and $+$ are the time ordered and anti-time ordered branches
of the contour, respectively.

Following the procedure outlined in Refs.~\cite{GogolinKomnikPRB06,ParkGalperin_FCS_PRB11},
at steady state we get the following expression for derivatives of
the cumulant generating function, $S=-i(t_f-t_i)$ $\mathcal{U}$ 
(here $\mathcal{U}$ is the adiabatic potential), in the counting fields
$\lambda_K^M$ ($M=P,E$)
\begin{equation}
\label{dUdlam}
\frac{\partial}{\partial\lambda_K^M} \mathcal{U}(\lambda_K^P,\lambda_K^E)
=-\int\frac{dE}{2\pi}\, O_M\, I_K^\lambda(E)
\end{equation}
where $O_M=1\, (E)$ for $M=P\, (E)$, and
\begin{align}
\label{defI}
 I_K^\lambda(E)\equiv&
  \mbox{Tr}\left\{
   \mathbf{\Sigma}_{K}^{<}(E)e^{i(\lambda_K^P+E\,\lambda_K^E)} \mathbf{G}_{\lambda}^{>}(E)
   \right. \\ & \left. \quad
  -\mathbf{G}_{\lambda}^{<}(E)\mathbf{\Sigma}_{K}^{>}(E)e^{-i(\lambda_K^P+E\,\lambda_K^E)}
  \right\}
  \nonumber
\end{align}
is the energy resolved dressed particle current at interface $K$, 
$\mbox{Tr}\{\ldots\}$ is the trace over the system subspace,
and $G_\lambda^{<(>)}$ is the lesser (greater) projections of the Green function 
obtained from a counting field dressed version of the Dyson equation 
(see e.g. Ref.~\cite{ParkGalperin_FCS_PRB11} for details).

While the expression for the derivatives of the adiabatic potential in the counting fields
can be easily formulated in terms of the field-dressed Green functions and self-energies
(see Eq.~(\ref{dUdlam}) above or Ref.~\cite{ParkGalperin_FCS_PRB11} 
for a detailed discussion), 
the corresponding expression for the adiabatic potential itself is more complicated. 
An explicit expression for the adiabatic potential within the NEGF based particle 
FCS for a single noninteracting level was derived in Ref.~\cite{GogolinKomnikPRB06}.
Exact results for the particle FCS for one-dimensional tight-binding 
junction models were presented in Ref~\cite{SchonhammerJPCM09}.
Here we consider the case of a multilevel noninteracting system. 
In particular, we show that for a single level coupled to its reservoirs 
(and possibly also to other levels with the latter not coupled to reservoirs) 
or for a multilevel system coupled to reservoirs through single molecular orbitals, 
the explicit expression for the adiabatic potential in the presence of both particle 
and energy counting fields is (see Appendix for details) 
\begin{align}
\label{Unonint}
& \mathcal{U}\left(\{\lambda\}\right)= i\int\frac{dE}{2\pi}\ln\bigg(1+\mathcal{T}(E)
 \nonumber\\ &\times
 \left\{
  f_L(E)[1-f_R(E)][e^{+i\left(\lambda_L^P-\lambda_R^P+E(\lambda_L^E-\lambda_R^E)\right)}-1]
  \right.\\ &\left.\ \
 +f_R(E)[1-f_L(E)][e^{-i\left(\lambda_L^P-\lambda_R^P+E(\lambda_L^E-\lambda_R^E)\right)}-1]
 \right\}
 \bigg)
 \nonumber
\end{align}
where
\begin{equation}
\label{TLandauer}
\mathcal{T}(E)\equiv \mbox{Tr}\{\Gamma_L(E)\, G^r(E)\,\Gamma_R(E)\, G^a(E)\}
\end{equation}
is the Landauer transmission coefficient at energy $E$.
Here $G^{r(a)}(E)$ are the retarded (advanced) projections of the system Green function
in absence of the counting fields, and $\Gamma_K(E)$ is the electron dissipation 
matrix at energy $E$ due to coupling to contact $K$ ($K=L,R$).
The size of the matrix is that of the molecular subspace of the problem. 
Below we consider systems for which expression (\ref{Unonint}) is satisfied.
In these systems, the electron dissipation rate matrices are always diagonal 
in the local basis. We denote by $\Gamma_L$ and $\Gamma_R$ the parameters 
characterizing the electron escape rates into the left and right contact, respectively. 

The particle and energy average currents and noises can be directly 
obtained from the adiabatic potential $\mathcal{U}$, Eq.~(\ref{Unonint}), as
\begin{align}
\label{defI}
&I^M_K = -\partial_{\lambda_K^M} \mathcal{U} \rvert_{\{\lambda\}=0} 
\\
\label{defStot}
&S^{M_1M_2}_{K_1K_2} = i\, \partial_{\lambda_{K_1}^{M_1}} \partial_{\lambda_{K_2}^{M_2}} \mathcal{U} \rvert_{\{\lambda\}=0},
\end{align}
where $K=L,R$ and $M=P,E$.
Explicitly, the average currents read
\begin{align}
\label{defIM}
I^M &\equiv I_L^M = -I_R^M \\
&= \int\frac{dE}{2\pi}\, O_M\, \mathcal{T}(E)\left[f_L(E)-f_R(E)\right] \nonumber
\end{align}
while the noises read
\begin{align}
\label{defSMM}
S^{M_1M_2} &\equiv S_{LL}^{M_1M_2}=S_{RR}^{M_1M_2}=-S_{LR}^{M_1M_2}=-S_{RL}^{M_1M_2} \\
&= S^{M_1M_2}_{shot} + S^{M_1M_2}_{therm} \nonumber ,
\end{align}
where the shot and the thermal (equilibrium) noise respectively read
\begin{align}
\label{defSs}
S^{M_1M_2}_{shot} =& \int\frac{dE}{2\pi}\,O_{M_1}\,O_{M_2}\, \\ 
& \times \mathcal{T}(E)\,\left(1-\mathcal{T}(E)\right) \left[f_L(E)-f_R(E)\right]^2 \nonumber \\
\label{defSt}
S^{M_1M_2}_{therm}=& \sum_{K=L,R}\int\frac{dE}{2\pi}\, O_{M_1}\,O_{M_2}\, \\ 
& \times \mathcal{T}(E)\,f_K(E)\left[1-f_K(E)\right] \nonumber .
\end{align}
Expressions (\ref{defI})-(\ref{defSt}) are exact for any non-interacting system bi-linearly coupled to two contacts.

\section{Efficiency fluctuation}\label{efffluct}

In order to operate as a thermoelectric junction the small quantum system 
is embedded between two leads $L$ and $R$ with $T_L<T_R$ and $\mu_L>\mu_R$ 
(see Fig.~\ref{fig1}b). 
The macroscopic efficiency of such a junction is defined as the ratio between 
the average power generated by the device, $\dot{W}=(\mu_L-\mu_R) I^P$, and 
the average heat taken from the hot reservoir which fuels the device, 
$\dot{Q}=-(I^E-\mu_R I^P)$, namely $\bar{\eta}=\dot{W}/\dot{Q}$.
It is upper bounded by the Carnot efficiency $\bar{\eta} \leq 1-T_L/T_R$.
The fluctuating efficiency on the other hand is defined as the ratio between 
the fluctuation power $w/t$ and heat flow $q/t$ measured at the level of a 
single experiential realization of duration $t$, namely $\eta=w/q$.
Efficiency fluctuations are not bounded and are characterized by the rate $J(\eta)$ 
at which the probability to observe a given efficiency $\eta$ decays during a long 
measurement realization \cite{EspositoVanDenBroeckNatCommun14, VanDenBroeckEspositoPRE14}
\begin{equation}
P(\eta) \stackrel{t \to \infty}{=} \exp{\{- J(\eta) t\}}.
\end{equation}
This rate is called the large deviation function (LDF) of efficiency.
It can be derived from the heat and work FCS obtained from 
the energy and heat FCS (\ref{Unonint}) as follows.
The heat entering the system from the hot (cold) reservoir is given by 
the right (left) energy current minus $\mu_R$ ($\mu_L$) times the right 
(left) particle current. 
At steady-state, the particle and energy currents are 
equal (but with opposite signs) at the two interfaces.
Therefore, by the first law of thermodynamics, the work generated 
by the particles moving across the system is equal to the sum of 
the heat from the left and right reservoir which is thus $\mu_R-\mu_L$ 
multiplied by the right particle current. 
This means that if $\lambda_Q$ counts the heat from the hot reservoir
and if $\lambda_W$ counts the work, we get that the heat and work FCS reads 
\begin{align}
\label{GFWorkHeat}
 & \mathcal{U}= i\int\frac{dE}{2\pi}\ln\bigg(1+T(E)
 \\ &\times
 \left\{
  f_L(E)[1-f_R(E)][e^{-i\left([E-\mu_R]\lambda_Q-[\mu_L-\mu_R]\lambda_W\right)}-1]
  \right.\nonumber\\ &\left.\ \ 
 +f_R(E)[1-f_L(E)][e^{+i\left([E-\mu_R]\lambda_Q-[\mu_L-\mu_R]\lambda_W\right)}-1]
 \right\}
 \bigg)
 \nonumber .
\end{align}
Introducing the slightly modified version of the adiabatic potential, 
$\phi\equiv -i\mathcal{U}$, and redefining the counting fields as 
$\gamma \equiv i\lambda_W$ and $\lambda\equiv i\lambda_Q$, we get that
\begin{align}
\label{defphi}
 &\phi(\gamma,\lambda)= \int\frac{dE}{2\pi}\ln\bigg(1+T(E)
 \nonumber\\ &\times
 \left\{
  f_L(E)[1-f_R(E)][e^{-\left([E-\mu_R]\lambda-[\mu_L-\mu_R]\gamma\right)}-1]
  \right.\\ &\left.\ \
 +f_R(E)[1-f_L(E)][e^{+\left([E-\mu_R]\lambda-[\mu_L-\mu_R]\gamma\right)}-1]
 \right\}
 \bigg).
 \nonumber
\end{align}
Note that the fluctuation theorem symmetry holds 
\begin{equation}
\label{FT}
 \phi\bigg(\gamma,\lambda\bigg)=
 \phi\bigg(-\frac{1}{T_L}-\gamma,\frac{1}{T_R}-\frac{1}{T_L}-\lambda\bigg)
\end{equation}
as can be verified using the property
\begin{align}
& f_R(E)[1-f_L(E)]e^{\left([E-\mu_R](\frac{1}{T_R}-\frac{1}{T_L}-\lambda) 
-[\mu_L-\mu_R](-\frac{1}{T_L}-\gamma)\right)} \nonumber \\
& = f_R(E)[1-f_L(E)]e^{-\frac{E-\mu_L}{T_L}+\frac{E-\mu_R}{T_R}}\, 
 e^{-\left([E-\mu_R]\lambda-[\mu_L-\mu_R]\gamma\right)} \\
& \equiv f_L(E)[1-f_R(E)]e^{-\left([E-\mu_R]\lambda-[\mu_L-\mu_R]\gamma\right)}.\nonumber
\end{align}
The efficiency LDF is finally obtain by setting $\lambda=\eta\,\gamma$ 
and minimizing $\phi$ relative to the field $\gamma$, namely
\cite{EspositoVanDenBroeckNatCommun14, VanDenBroeckEspositoPRE14}
\begin{equation}
\label{defJ}
J(\eta) = -\min_\gamma \phi(\gamma,\eta\gamma).
\end{equation}
The convexity of (\ref{defphi}) together with the fluctuation theorem symmetry (\ref{FT}) 
has been used in classical systems to prove two important results. First, the single minimum 
in $J(\eta)$ (i.e. the most probable efficiency) corresponds to the macroscopic efficiency 
$\bar{\eta}$, second, the single maximum in $J(\eta)$ (i.e. the least likely efficiency) 
corresponds to the Carnot efficiency $1-T_L/T_R$ \cite{EspositoVanDenBroeckNatCommun14, 
VanDenBroeckEspositoPRE14}. By showing that the fluctuation theorem symmetry (\ref{FT})
holds for the adiabatic potential of quantum junctions, we thus generalized these 
remarkable results to the quantum realm.

In the limit of weak system-lead coupling, $\Gamma \equiv \Gamma_L+\Gamma_R\to 0$,
Eq.~(\ref{defphi}) reduces to the QME approach prediction~\cite{EspositoHarbola07, EspositoRMP09}
\begin{align}
\label{defphiQME}
&\phi(\gamma,\lambda)= \\
& \sum_s \bigg(-\frac{\Gamma_s(E_s)}{2}+\bigg[\left(\frac{\Gamma_s(E_s)}{2}\right)^2 +\Gamma^L_s(E_s)\Gamma^R_s(E_s) \nonumber \\ 
& \times \left\{ f_L(E_s)[1-f_R(E_s)] [e^{-\left([E_s-\mu_R]\lambda-[\mu_L-\mu_R]\gamma\right)}-1]  \right. \nonumber \\  
& \left. +f_R(E_s)[1-f_L(E_s)][e^{+\left([E_s-\mu_R]\lambda-[\mu_L-\mu_R]\gamma\right)}-1] \right\} \bigg]^{1/2} \bigg) \nonumber .
\end{align}
Here $\sum_s\ldots$ is the sum over the eigenorbitals of the system with eigenenergies $E_s$, 
and $\Gamma_s(E_s) \equiv \Gamma^L_s(E_s)+\Gamma^R_s(E_s)$ is the total escape rate from 
the eigenorbital $s$ evaluated at energy of the the orbital.
The quasi-classical nature of this result is manifest since Eq.~(\ref{defphiQME}) disregards 
the reservoir induced correlations between the eigenorbitals of the system. 
This form of adiabatic potential was used in Refs.~\cite{EspositoVanDenBroeckNatCommun14,
VanDenBroeckEspositoPRE14} together with (\ref{defJ}) to calculate efficiency fluctuations 
in a photoelectric device.

\begin{figure}[t]
\centering\includegraphics[width=\linewidth]{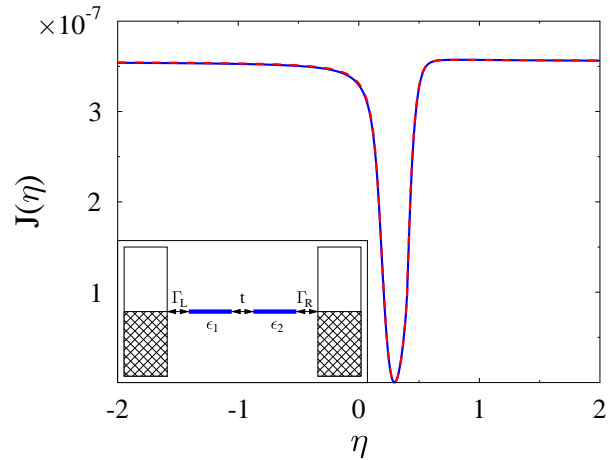}
\caption{\label{fig2}
(Color online) Efficiency LDF for a two-level bridge calculated within the NEGF 
(solid line, blue) and the QME (dashed line, red) approaches. See text for parameters.
}
\end{figure}

\section{Numerical examples}\label{numres}

We now compare the efficiency fluctuations (\ref{defJ}) predicted using the NEGF 
heat and work FCS (\ref{defphi}) with the QME prediction (\ref{defphiQME}). 
Since we exclusively consider non-interacting models, we emphasize that the NEGF 
treatment is exact while the QME approach is an approximate approach only valid 
in the weak coupling limit to the contact and which neglects coherences between 
system eigenstates (we use the rotating wave approximation to guarantee positivity). 
The discrepancies between these two approaches can thus be attributed to broadening 
effects induced by strong coupling and to eigenbasis coherences. 

The calculations are performed by numerically evaluating the adiabatic potential 
$\phi(\gamma,\eta \gamma)$ (using Eq.(\ref{defphi}) for the NEGF and Eq.(\ref{defphiQME}) 
for the QME) and numerically minimizing it as a function of the counting field 
$\gamma$ for a fixed value of the efficiency $\eta$ according to Eq.(\ref{defJ}).

Unless specified otherwise, the parameters of the calculations are 
$T_L=100$~K, $T_R=600~K$, $\mu_L=0.02$~eV and $\mu_R=0$. 
We use the wide band approximation which assumes that the electron escape 
rates $\Gamma_L$ and $\Gamma_R$ are energy-independent constants. 
The NEGF calculations were performed on an energy grid
spanning the region from $-1$ to $1$~eV with step $10^{-5}$~eV.

We start by considering the two-level bridge model depicted in inset in 
Fig.~\ref{fig2} when the system is weakly coupled to the contacts.
The position of the levels is $\varepsilon_1=\varepsilon_2=0.1$~eV, 
the electron hopping parameter is $t=0.05$~eV, and the electron escape 
rates are $\Gamma_L=\Gamma_R=2\cdot10^{-4}$~eV. As expected, in this 
regime both the NEGF and the QME predictions for the efficiency
fluctuation coincide (compare the solid and the dashed lines 
in Fig.~\ref{fig2}). 
Large values of $J(\eta)$ indicates unlikely efficiency fluctuations 
while the minimum is the most likely efficiency $\bar{\eta}$ corresponding 
to the macroscopic efficiency considered in traditional thermodynamics.
Although hardly seen on this figure, the most unlikely efficiency 
is located at the Carnot efficiency $1-T_L/T_R \approx 0.83$.
The probability distribution in this regime is thus quite 
narrowly centered around the most likely efficiency.  

\begin{figure}[t]
\centering\includegraphics[width=\linewidth]{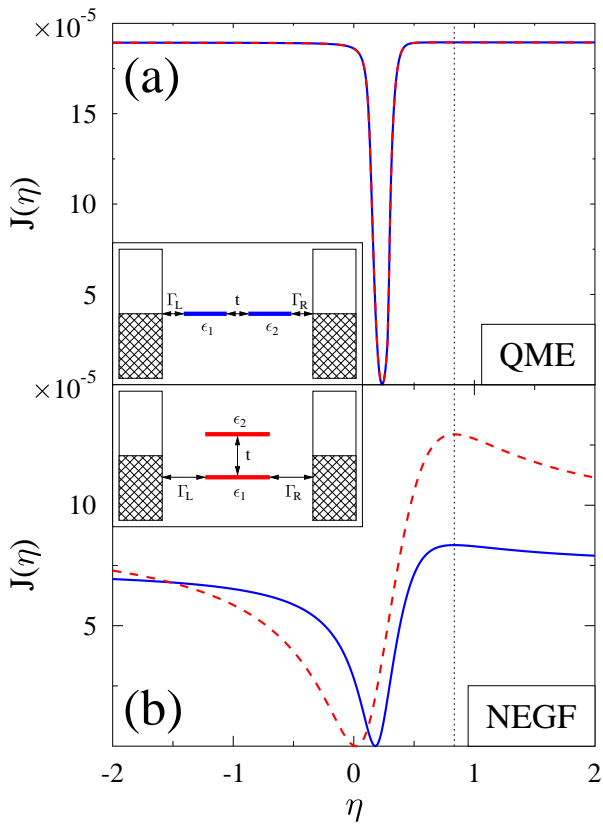}
\caption{\label{fig3}
(Color online) Efficiency LDF for a two-level bridge (top inset; solid line, blue) 
and a single level coupled to an isolated orbital (bottom inset; dashed line, red),
calculated within the (a) QME and (b) NEGF approaches. 
The vertical dashed line shows the Carnot efficiency. See text for parameters.
}
\end{figure}

We consider two types of junctions, a two-level bridge (top inset in Fig.~\ref{fig3})
and a single level junction coupled to an isolated orbital (bottom inset in Fig.~\ref{fig3}).
Both junctions are in regimes where the system is strongly coupled to the contacts.
The latter is the simplest model often used to describe destructive interference 
effect in transport through a junction (see e.g. Ref.~\cite{EspGalpJPCC10}).
The position of the levels is $\varepsilon_1=\varepsilon_2=0.12$~eV, 
the electron hopping parameter is $t=0.05$~eV, and the electron 
escape rates are $\Gamma_L=\Gamma_R=0.1$~eV. 
Figure~\ref{fig3}a shows that QME results of the two models are identical. 
This result stems from the fact that in the rotating wave approximation, the 
QME neglects coherences in the system eigenbasis \cite{EspGalpJPCC10, EspGalpPRB09}.
Fig.~\ref{fig3}b shows the exact efficiency fluctuations for the two models.
The interference effects responsible for the discrepancy between the two curves 
do not significantly alter the qualitative shape of the efficiency LDF. 
However, when comparing Figs.~\ref{fig3}a and b, we note that 
the broadening effects resulting from the strong coupling to the contacts 
clearly tend to increase the magnitude of the efficiency fluctuations 
and also intensifies the asymmetry of the fluctuations around the most 
likely value. We note that even the most likely value is affected.
The least likely value is nevertheless still exactly located at the Carnot efficiency.

\begin{figure}[t]
\centering\includegraphics[width=\linewidth]{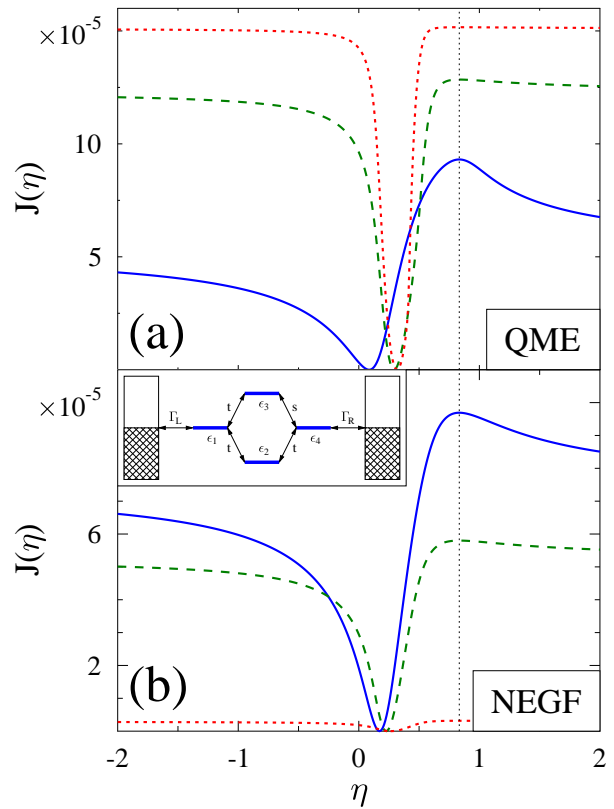}
\caption{\label{fig4}
(Color online) Efficiency LDF for a donor-bridge-acceptor junction
calculated within the (a) QME and (b) NEGF approaches. 
Results are shown for constructive interference ($s=t$; solid line, blue),
single path ($s=0$; dashed line, green), and destructive interference 
($s=-0.8\, t$; dotted line, red). The vertical dashed line shows 
the Carnot efficiency. Other parameters are as in Fig.~\ref{fig3}.
}
\end{figure}

We now turn to the donor-bridge-acceptor (DBA) junction depicted in the inset of Fig.~\ref{fig4}. 
This setup enables to study the effect of intra-molecular interference on efficiency fluctuations. 
We see that the trend predicted by the QME, when moving from constructive to destructive interference 
(solid to dashed to dotted line), is the opposite of the real trend obtained using the exact NEGF. 
It is interesting to observe that destructive interference tend to increase the most likely 
efficiency but at the same time significantly increase the magnitude of the efficiency fluctuations. 
In other words, the performance of the junction increases but at the cost of becoming less reproducible. 

\begin{figure}[t]
\centering\includegraphics[width=\linewidth]{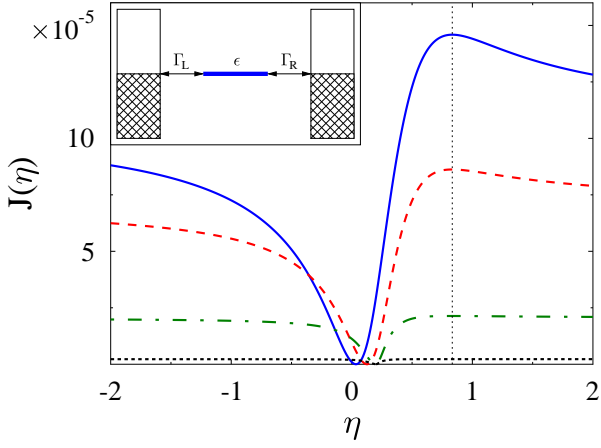}
\caption{\label{fig5}
(Color online) Efficiency LDF calculated within the NEGF for a single level junction. 
The results are shown for several level-contacts coupling strengths
ranging from the strongest (solid line, blue) to the weakest (dotted line, black). 
The vertical dashed line shows the Carnot efficiency. See text for parameters.
}
\end{figure}

As a final example we consider a single level junction (see inset in Fig.~\ref{fig5}). 
Within the QME approach, the efficiency does not fluctuate in this model because  heat 
and work are directly proportional to each other, a condition known as tight coupling \cite{EspoPoleVerl14}. 
However, the NEGF approach breaks the tight coupling condition due to the hybridization 
of the molecular level with the states in the contacts. 
The position of the level is taken as $\varepsilon=0.1$~eV and Figure~\ref{fig5} shows 
the results of calculations for several strengths of the system-reservoir coupling: 
$\Gamma_L=\Gamma_R=0.1$~eV (solid line), $0.05$~eV (dashed line), $0.01$~eV (dash-dotted line), 
and $0.001$~eV (dotted line). As $\Gamma \to 0$ (weak coupling limit) the distribution 
becomes very narrow and centered around the macroscopic efficiency $(\mu_L-\mu_R)/(\varepsilon-\mu_R)$. 

We now discuss ways to relate the efficiency LDF to experimentally measurable characteristics of the junction.
For the setup sketched in Fig.~\ref{fig1}b, the average power and the heat flux from the hot reservoir are 
\begin{align}
\dot{W} =& \Delta\mu\, I^P \\
\dot{Q} =& -\left(I^E - \mu_R\, I^P\right),
\end{align}
where $I^P$ and $I^E$ are defined in Eq.~(\ref{defIM})
and $\Delta \mu \equiv \mu_L-\mu_R$. 
In the linear response regime (obtained by linearizing the Fermi distributions in $1/T_{L(R)}$ and 
$\mu_{L(R)}/T_{L(R)}$ around equilibrium $\mu_L=\mu_R=E_F$ and $T_L=T_R=T$), we get that
\begin{align}
&\dot{W} \approx G\,\Delta\mu^2 +L\,\Delta\mu\,\Delta\beta \\
&\dot{Q} \approx R\,\Delta\mu+F\,\Delta\beta
\end{align}
where $\Delta\beta=1/T_L-1/T_R$ and
\begin{align}
\label{defcond}
G =& -\int\frac{dE}{2\pi}\, \mathcal{T}(E)\, f'(E)\, \frac{1}{T_L} \\
L =& \int\frac{dE}{2\pi}\, \mathcal{T}(E)\, f'(E)\, (E-\mu_R) \\
R =& \int\frac{dE}{2\pi}\,  \mathcal{T}(E)\, f'(E)\, \frac{E-\mu_R}{T_L} \\
F =& -\int\frac{dE}{2\pi} \, \mathcal{T}(E)\, f'(E)\, (E-\mu_R)^2.
\end{align}
Here $f'(E)=[d/dx\, 1/(e^x+1)]_{x=(E-E_F)/T}$ and $R=L/T_L$.
The coefficients in (\ref{defcond}) are related to experimentally measurable
quantities. 
Indeed, $G$ is the electrical conductance, and if $\kappa$ denotes the heat 
conductance and $S$ the Seebeck coefficient, we have that
\begin{equation}
\kappa = \frac{F}{T_L\, T_R} 
; \qquad
S = \frac{L}{G\, T_L\, T_R}.
\end{equation}
Thus following Ref.~\cite{EspositoVanDenBroeckNatCommun14}, 
in the linear response regime the efficiency LDF can be expressed in terms of 
these measurable quantities as
\begin{equation}
\label{Jlinresp}
J(\eta) =\frac{\left[\eta (\kappa\,\Delta T+G\,S\,T_R\,\Delta\mu)
+G\, S\, \Delta T\Delta\mu +G\,\Delta\mu^2 \right]^2}
{4\left[\eta^2\, \kappa\, T_L\, T_R
+2\, \eta\, G\, S\, T_L\, T_R\, \Delta\mu
+G\, T_L\,\Delta\mu^2\right]
},
\end{equation}
where $\Delta T=T_R-T_L$.  

\begin{figure}[t]
\centering\includegraphics[width=\linewidth]{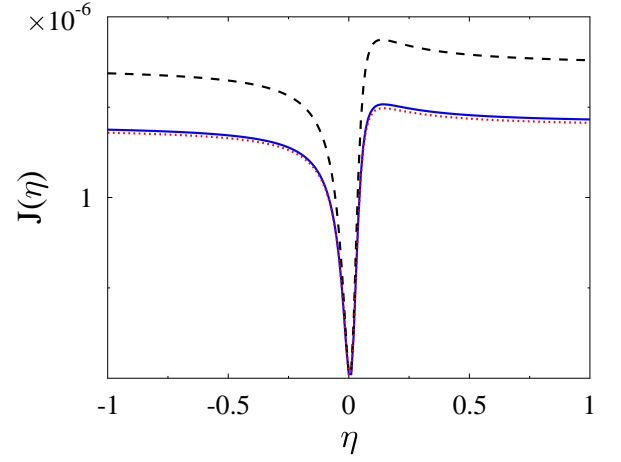}
\caption{\label{fig6}
(Color online) Efficiency LDF for a single level junction (see inset in Fig.~\ref{fig5}) 
calculated for experimentally relevant parameters.
The predictions of the exact NEGF calculations (solid line, blue)
are compared to the linear response predictions (\ref{Jlinresp}) (dashed line, black), 
and to the Gaussian approximation predictions (\ref{Jneq}) (dotted line, red). 
See text for parameters.
}
\end{figure}

We now attempt to estimate the efficiency LDF beyond the linear regime, solely 
in terms of the particle and energy nonequilibrium currents and noises, 
Eqs.~(\ref{defIM})-(\ref{defSt}).
Note that in molecular junctions, the particle and energy currents as well as the particle noise 
are experimentally measurable~\cite{RuitenbeekNL06,TalRuitenbeekPRL08,Tal2008,BerndtPRL12} and the energy noise
will soon become measurable~\cite{BernardDoyonJPhysA12,HanggiPSSB13,SplettstroesserPRB14}.
To do so, we approximate the cumulant generating function (\ref{defphi}) 
by a quadratic expansion in counting fields $\gamma$ and $\lambda$ around
point $\gamma=\lambda=0$. This is a Gaussian assumption which leads to
\begin{equation}
\phi(\gamma,\eta\gamma) \approx a \gamma^2 + b \gamma,
\end{equation}
where we used the fact that $\phi(0,0)=0$, and defined
\begin{align}
a \equiv& \frac{\eta^2}{2} S^{EE} +\frac{\left(\mu_L-\mu_R[1-\eta]\right)^2}{2} S^{PP} \\ 
&- \eta \left(\mu_L-\mu_R[1-\eta]\right) S^{PE} \nonumber \\
b \equiv& -\eta\, I^E + \left(\mu_L-\mu_R[1-\eta]\right) I^P,
\end{align}
which are solely expressed in terms of the measurable nonequilibrium particle and 
energy fluxes, Eq.~(\ref{defIM}), and of the nonequilibrium noise 
characteristics of the junction (\ref{defSMM}).
Within this Gaussian approximation, we find that 
\begin{equation}
\label{Jneq}
J(\eta) = \frac{b^2}{4\, a}.
\end{equation}

We have thus shown that efficiency fluctuations are experimentally measurable
close to equilibrium (\ref{Jlinresp}) and in the Gaussian approximation (\ref{Jneq}). 
We now verify the validity of these approximations in Fig.~\ref{fig6} where
the exact efficiency LDF, Eqs.~(\ref{defphi}) and (\ref{defJ}), is compared to  
the linear response, Eq.~(\ref{Jlinresp}), and 
the Gaussian approximation, Eq.~(\ref{Jneq}), result for an experimentally relevant 
set of parameters: $T_L=300$~K, $T_R=350$~K, $\mu_L=0.002$~eV, $\mu_R=0$~\cite{MajumdarScience07}. 
We also set $\varepsilon=0.1$~eV and $\Gamma_L=\Gamma_R=0.1$~eV.
We see that near the minimum corresponding to the macroscopic efficiency, the three curves
coincide (thus justifying the use of linear response to estimates of average quantities). 
At the same time, the efficiency fluctuations are poorly captured by the linear 
response approximation (dashed line) but reproduced quite well by the Gaussian approximation (dotted line). 

\section{Conclusion}\label{conclude}

We studied the thermoelectric properties of nanoscale junctions. 
Since stochasticity and quantum coherence are expected to be important 
at small scale, we proposed to characterize the performance of these 
devices by studying efficiency fluctuations rather than the widely used 
figure of merit which is intrinsically restricted to the linear regime. 
We provided a systematic procedure to study efficiency fluctuations
which accounts for all quantum effects and is based on the work and 
heat FCS obtained within the NEGF formalism. As predicted for classical 
dynamics in Ref. \cite{EspositoVanDenBroeckNatCommun14}, the most likely 
efficiency coincides with the macroscopic efficiency, while the least likely 
efficiency corresponds to the Carnot efficiency.  
We used simple models with realistic molecular junction parameters to 
compare our NEGF based method to the commonly used QME approach. We showed
that the latter may fail qualitatively for strong system-reservoir coupling 
due to its inability to properly account for quantum coherences in the system. 
We finally proposed a method to estimate efficiency fluctuations using the 
experimentally measurable particle and energy nonequilibrium currents and noises. 
Linear response and Gaussian approximations were proposed as ways to
construct efficiency fluctuations from experimental measurements.
We showed that while linear response approach, often used in the experimental 
literature to discuss thermoelectric properties of junctions, captures the 
macroscopic efficiency, it fails to account for the efficiency fluctuations.
At the same time, the Gaussian approximation was shown to work 
very well within experimentally relevant range of parameters.

\begin{acknowledgments}
M.E. is supported by the National Research Fund, Luxembourg in the frame of project FNR/A11/02.
M.G. gratefully acknowledges support by the Department of Energy (Early Career Award, DE-SC0006422).
\end{acknowledgments}
\begin{widetext}

\appendix

\section{Cumulant generating function of a multilevel non-interacting system}

Here we derive the general form of the adiabatic potential $\mathcal{U}(\lambda)$ for a noninteracting $n$-level system. For simplicity, we consider the specific case of one 
particle counting field $\lambda$ in the left molecule-contact interface. 
Multiple counting fields and/or energy FCS are formulated similarly.
We first find expression for the potential derivative in the counting field,
Eqs.~(\ref{dUdlam}) and (\ref{defI}), and then integrate it in the field to get the 
potential itself.

We start by writing the dressed Green Function $G(\lambda)$ 
as a $2n \times 2n$ dimensional block matrix in the Keldysh contour
\begin{equation}
  G(\lambda) =
  \begin{bmatrix}
    G_\lambda^{c}&G_\lambda^{<}\\*[0.1cm]
    G_\lambda^{>}&G_\lambda^{\tilde c}
  \end{bmatrix},
\end{equation}
inverse of which is \cite{GogolinKomnikPRB06}
{\small
\begin{equation}
\label{eq:apInvG}
  G^{-1}(\lambda) =
  \begin{bmatrix}
    -i \Gamma_L ( f_L(E)- 1/2 )-i \Gamma_R (f_R(E) - 1/2) +IE - H_M & i e^{i \lambda} \Gamma_L f_L(E)+ i \Gamma_R f_R(E)\\
    - i e^{-i \lambda} \Gamma_L (1 - f_L(E))- i \Gamma_R(1- f_R(E))&-i\Gamma_L(f_L(E)-1/2)- i \Gamma_R(f_R(E)-1/2)-IE + H_M
  \end{bmatrix} .
\end{equation}
}
We will use Jacobi's formula for the derivative of the determinant of a matrix that in our case reads
\begin{equation}
\label{eq:apJac}
  \frac{d}{d \lambda}\det(G^{-1}(\lambda)) = {\rm Tr} \left\{ {\rm adj}\left(G^{-1}(\lambda)\right)\frac{d}{d\lambda}G^{-1}(\lambda)\right \} ,
\end{equation}
where ${\rm adj}(M)$ denotes the adjugate matrix of a matrix $M$ 
(${\rm adj}(M) M = I\det(M) = M\, {\rm adj}(M)$).  
For our consideration, it will be important to work with special submatrices of $G^{-1}$. 
For an $n \times n$ matrix $M$, we define $M(j|i)$ to be the $(n-1)\times (n-1)$ matrix that is obtained from $M$ by removing the $j$th row and the $i$th column. 
In this notation the $(i,j)$-matrix element for the adjugate of $M$ can be expressed as 
$ {\rm adj}(M)_{ij}= (-1)^{i+j} \det(M(j|i))$.
Also below $M[j_1 \dots j_r| i_1 \dots i_r]$ will denote the submatrix of $M$ composed of rows $j_1 \dots j_r$ and columns $i_1 \dots i_r$.

The first step in the derivation is to obtain from Eq.\ (\ref{eq:apInvG})
\begin{equation}
  \frac{d}{d\lambda}G^{-1}(\lambda)=
  \begin{bmatrix}
    0 & - e^{i \lambda} \Gamma_L f_L(E)\\
    - e^{-i \lambda} \Gamma_L (1 - f_L(E))& 0
  \end{bmatrix},
\end{equation}
and utilizing Eq.\ (\ref{eq:apJac}) calculate
\begin{align}
  \frac{1}{\det(G^{-1}(\lambda))}\frac{d}{d \lambda}\det(G^{-1}(\lambda)) =&\frac{1}{\det(G^{-1}(\lambda))} {\rm Tr} \left\{ {\rm adj}\left(G^{-1}(\lambda)\right)\frac{d}{d\lambda}G^{-1}(\lambda)\right \} \notag \\
=& {\rm Tr} \left\{ \frac{{\rm adj}\left(G^{-1}(\lambda)\right)}{\det(G^{-1}(\lambda))} 
\begin{bmatrix}
    0 & - e^{i \lambda} \Gamma_L f_L(E)\\
    - e^{-i \lambda} \Gamma_L (1 - f_L(E))& 0
  \end{bmatrix}
\right \} \notag \\
=& {\rm Tr} \left\{  
  \begin{bmatrix}
    G_\lambda^{c}(E)&G_\lambda^{<}(E)\\*[0.1cm]
    G_\lambda^{>}(E)&G_\lambda^{\tilde c}(E)
  \end{bmatrix}
\begin{bmatrix}
    0 & - e^{i \lambda} \Gamma_L f_L(E)\\
    - e^{-i \lambda} \Gamma_L (1 - f_L(E))& 0
  \end{bmatrix}
\right \} \notag \\
=&{\rm Tr} \left\{ G_\lambda^< (E) (-e^{-i \lambda}) \Gamma_L(1- f_L(E))+G_\lambda^> (E)(- e^{i \lambda})\Gamma_L f_L(E)\right\} \notag\\
=&i I_L^\lambda(E) .
\end{align}
Using this last result in Eq.(\ref{dUdlam}) and integrating with respect to the counting 
field $\lambda$ leads to
\begin{equation}
\label{eq:apUlam1}
\mathcal{U}(\lambda)=  i \int \frac{dE}{2 \pi} \ln \left[ \frac{\det(G^{-1}(\lambda))}{\det(G^{-1}(0))} \right].
\end{equation}
where we used the known property $\mathcal{U}(0)=0$.
Eq.~(\ref{eq:apUlam1}) is the first important result.

We now have to evaluate the determinants inside the logarithm in Eq.~(\ref{eq:apUlam1}).
The determinants can be evaluated after applying elementary transformations 
to $G^{-1}$. First, we notice that we can write
\begin{equation}
\label{eq:apGlam}
  \det (G^{-1}(\lambda)) =
\begin{vmatrix}
-\Sigma^>(E)+G^{a, -1} &-i (1 -e^{i\lambda}) \Gamma_L f_L(E) +\Sigma^<(E)\\
    i(1-e^{-i \lambda}) \Gamma_L (1 - f_L(E))+\Sigma^>(E) &-G^{r,-1}-\Sigma^>(E)
  \end{vmatrix},
\end{equation}
where $G^{r, -1}=I E - H_M + i (\Gamma_L + \Gamma_R)/2$ and $G^{a,-1}= (G^{r, -1})^\dagger$, by adding and subtracting to each submatrix in Eq.\ (\ref{eq:apInvG}) appropriate matrices. 
Then we add to the $i$-th row, $i \leq n$ the $(n+i)$th row of the matrix. After which, 
on the resulting matrix, we add the $(n+j)$-th column to the $j$-th column 
for each $j\leq n$. This leads to
\begin{equation}
\label{eq:apDetG1}
  \det (G^{-1}(\lambda))= \begin{vmatrix}
i (1-e^{-i \lambda})\Gamma_L (1-f_L(E))+G^{a, -1} & -i (1 -e^{i \lambda}) \Gamma_L f_L(E) +i (1-e^{-i \lambda})(1 -f_L(E))\Gamma_L\\
    i(1-e^{-i \lambda}) \Gamma_L (1 - f_L(E))+\Sigma^>(E)&i(1- e^{-i \lambda})\Gamma_L (1-f_L(E))-G^{r,-1}
  \end{vmatrix}
\end{equation}
Setting $\lambda=0$ we arrive at the result for the first of the determinants in
Eq.~(\ref{eq:apUlam1})
\begin{equation}
\label{eq:apGlam0}
\det(G^{-1}(0))=\det(G^{a,-1})\det(-G^{r, -1}),
\end{equation}
  
To get the second determinant in Eq.~(\ref{eq:apUlam1}) we have to work with
the general form of Eq.~(\ref{eq:apGlam}). 
Explicit evaluations lead to an expression that can be grouped in powers of $(1-e^{-i \lambda})$ and $(1-e^{i \lambda})$ of at most $n$ power. 
In particular, noticing that 
$(1 - e^{-i \lambda})(1-e^{i \lambda})= (1 - e^{-i \lambda})+(1 - e^{i \lambda})$ 
we can write
\begin{equation}
\label{eq:apGlamPol}
  \det(G^{-1}(\lambda))= \det(G^{a, -1}) \det(-G^{r,-1})+ \sum_{s=1}^n(1 - e^{-i \lambda})^s a_s +(1 - e^{i \lambda})^s b_s , 
\end{equation}
where $a_s$ and $b_s$ are the coefficients of the polynomial given by \cite{MarcusMinc_1988}
\begin{align}
  a_s=&(i)^s (1- f_L(E))^s \sum_{\alpha \in Q_{s,n}} \sum_{\beta \in Q_{s,n}} (-1)^{\sigma(\alpha+n)+\sigma(\beta)} \det(\Gamma_L[\alpha| \beta]) \det(N(\alpha+n|\beta)) \label{eq:apas}\\
  b_s=&(-i)^s(f_L(E))^s \sum_{\alpha \in Q_{s,n}} \sum_{\beta \in Q_{s,n}} (-1)^{\sigma(\alpha)+\sigma(\beta+n)}  \det (\Gamma_L[\alpha| \beta]) \det(M(\alpha|\beta+n)) \label{eq:apbs}
\end{align}
where $N$ and $M$ are $2n \times 2n$ matrices given by
\begin{equation}
  N=
\begin{bmatrix}
-\Sigma^>(E)+G^{a, -1}& i \Gamma_R f_R(E)\\
\Sigma^>(E)& -G^{r, -1}-\Sigma^>(E)
\end{bmatrix},
\qquad
  M=
\begin{bmatrix}
-\Sigma^>(E)+G^{a, -1}& \Sigma^<(E)\\
-i \Gamma_R (1 - f_R(E))& -G^{r, -1}-\Sigma^>(E)
\end{bmatrix},
\end{equation}
$Q_{s,n}$ is the set of s-tuples $(i_1, \dots, i_s)$ of natural numbers 
such that $1\leq i_1 < i_2< \dots < i_{s-1} < i_s \leq n$, $\sigma(\alpha)= \sum \alpha_i$ 
for $\alpha \in Q_{s,n}$, and $\alpha + n=(\alpha_1+n, \alpha_2+n, \dots, \alpha_s+n)$. 
From equations (\ref{eq:apas}) and (\ref{eq:apbs}) we have in particular 
$a_n=((1-f_L(E))f_R(E))^n \det(\Gamma_L \Gamma_R)$ and 
$b_n= ((1-f_R(E))f_L(E))^n \det(\Gamma_L \Gamma_R)$.
Eqs.~(\ref{eq:apGlamPol})-(\ref{eq:apbs}) give the most general form for the
second determinant in Eq.~(\ref{eq:apUlam1}).

Finally, substituting Eqs.~(\ref{eq:apGlam0}) and (\ref{eq:apGlamPol}) into 
(\ref{eq:apUlam1}) we obtain the general form for the adiabatic potential
\begin{equation}
\label{eq:apUgen}
 \mathcal{U}(\lambda)= i\int \frac{dE}{2 \pi} \ln \left[ 1+ \sum_{s=1}^n \frac{a_s \,(1 - e^{-i \lambda})^s}{\det(G^{a, -1}) \det(-G^{r,-1})} + \frac{b_s \,(1 - e^{i \lambda})^s}{\det(G^{a, -1}) \det(-G^{r,-1})}\right], 
\end{equation}
with the coeffcients $a_s$ and $b_s$ given by Eqs.\ (\ref{eq:apas}) and (\ref{eq:apbs}). 

We now consider two specific examples where we can recover 
the expression derived in Ref.~\cite{GogolinKomnikPRB06}  for a single level junction
from our general result, Eq. (\ref{eq:apUgen}).
\begin{enumerate}
\item {\bf One level}.\\ By direct computation of Eqs.\ (\ref{eq:apas}) and (\ref{eq:apbs}) we find
  \begin{align*}
    a_1&=\Gamma_L \Gamma_R (1 - f_L(E)) f_R(E)&&b_1=\Gamma_L \Gamma_R (1- f_R(E))f_L(E).
  \end{align*}
Thus
\begin{align*}
  \frac{\det(G^{-1}(\lambda))}{\det(G^{-1}(0))}=& 1+ \frac{\Gamma_R \Gamma_L}{G^{a, -1}(- G^{r,-1})}\left((1 - e^{-i \lambda})(1-f_L(E))f_R(E) + (1 - e^{i \lambda})(1-f_R(E))f_L(E) \right)\\
=& 1+ G^r(E) \Gamma_R G^a(E) \Gamma_L \left((e^{i \lambda}-1)(1-f_L(E))f_R(E)+(e^{-i \lambda}-1)(1-f_L(E))f_R(E) \right)\\ 
\end{align*}
which yields the expression for adiabatic potential derived in 
Ref.~\cite{GogolinKomnikPRB06}.

\item {\bf n-level system coupled to the contacts through single orbitals.}\\
 In this case $\Gamma_L$ and $\Gamma_R$ are $n \times n $ matrices with all entries 
 equal zero but one element in the diagonal. Examples of systems of this kind are 
 the two level bridge (see inset in Fig.~\ref{fig2}) or D-B-A type of the junction 
 (see inset in Fig.~\ref{fig4}). 
 Here we can take $[\Gamma_L]_{ij}=\delta_{1j}\delta_{i1} \gamma_L$ 
 and $[\Gamma_R]_{ij}=\delta_{nj}\delta_{in} \gamma_R$, 
 which results in $a_s=b_s=0$ for $s>1$ and
 \begin{align}
   \label{eq:apa1}
   a_1=& i (1- f_L(E)) (-1)^{1+n+1} \det(\Gamma_L[1|1]) \det(N(1+n|1)) \notag\\
   =& i (1- f_L(E)) (-1)^{n} \gamma_L \begin{vmatrix}
-\Sigma^>(|1)(E)+G^{a, -1}(|1) &i\Gamma_R f_R(E)\\
    -i\Gamma_R(1|1) (1 - f_R(E)) &-G^{r,-1}(1|)-\Sigma^>(1|)(E)
  \end{vmatrix} \notag\\
 =& i (1- f_L(E)) (-1)^{n} \gamma_L (-1)^{n+2n-1}  i \gamma_R f_R(E)
\begin{vmatrix}
-\Sigma^{>}(n|1)(E)+G^{a, -1}(n|1) &0\\
    -i\Gamma_R(1|1) (1 - f_R(E)) &-G^{r,-1}(1|n)-\Sigma^>(1|n)(E)
  \end{vmatrix}  \notag\\
 =& (1- f_L(E)) f_R(E) \gamma_L \gamma_R \det(G^{a, -1}(n|1)) \det(-G^{r,-1}(1|n))
\end{align}
where we used $\Sigma^>(n|1)(E)=\Sigma^>(1|n)(E)=0$. 
Similarly
\begin{equation}
  \label{eq:apb1}
 b_1= f_L(E) (1- f_R(E)) \gamma_L  \gamma_R \det(G^{a, -1}(1|n)) \det(-G^{r,-1}(n|1)).\\
\end{equation}
From the definition of the adjugate we have  $(-1)^{1+n}\det(G^{a, -1}(1|n)) = {\rm adj}(G^{a,-1})_{n1}$, $(-1)^{n+1}\det(-G^{r, -1}(n|1)) = {\rm adj}(-G^{r,-1})_{1n}$. 
Also in this particular case ${\rm adj}(G^{a,-1})_{1 n}= {\rm adj}(G^{r,-1})_{1 n}$ and ${\rm adj}(G^{a,-1})_{n 1}= {\rm adj}(G^{r,-1})_{n 1}$. 
Finally, substituting Eqs.\ (\ref{eq:apa1}) and (\ref{eq:apb1}) into Eq.\ (\ref{eq:apUgen}) and rearranging terms, we get
\begin{align}
\label{eq:apLev}
 \mathcal{U}(\lambda)=& i \int \frac{dE}{2 \pi}\ln \bigg[ 1+ {\rm Tr}\{G^r(E) \Gamma_R G^a(E) \Gamma_L\} 
 \\ &\times
 \left((e^{i \lambda}-1)(1-f_L(E))f_R(E)+(e^{-i \lambda}-1)(1-f_L(E))f_R(E) \right) \bigg] 
 \nonumber
\end{align}
Eq.\ (\ref{eq:apLev}) is the Levitov-Lesovik formula for a multi-level system.
\end{enumerate}
\end{widetext}


\end{document}